\documentclass[aps,pre,amsmath,amssym,amsthm,superscriptaddress,reprint, nofootinbib]{revtex4-2}
\usepackage[colorlinks=true,urlcolor=blue,citecolor=blue,linkcolor=blue]{hyperref}

\usepackage{graphicx}
\usepackage{dcolumn}
\usepackage{kotex}
\usepackage{natbib}
\usepackage{color}
\usepackage{amsfonts}
\usepackage{hyperref}
\usepackage{algorithm}
\usepackage{algcompatible}
\usepackage{newfloat}
\usepackage{makecell}
\usepackage{bm}

\usepackage{setspace}
\usepackage{multirow}
\usepackage{sidecap}
\usepackage{enumerate}

\usepackage{orcidlink}

\definecolor{mygray}{gray}{0.5}

\begin{document}

\title{Impact of capacity volatility and input substitutability on supply chain resilience}

\author{Jaeseok Hur\orcidlink{0009-0007-1976-275X}}
\affiliation{Department of Physics, Korea Advanced Institute of Science and Technology, Daejeon 34141, Korea}

\author{Juha Jang}
\affiliation{Department of Physics, Korea Advanced Institute of Science and Technology, Daejeon 34141, Korea}

\author{Meesoon Ha\orcidlink{0000-0003-1279-5325}}
\email[Contact author: ]{msha@chosun.ac.kr}
\affiliation{Department of Physics Education, Chosun University, Jeonam-Gwangju, 61452, Korea}

\author{Hawoong Jeong\orcidlink{0000-0002-2491-8620}}
\email[Contact author: ]{hjeong@kaist.edu}
\affiliation{Department of Physics, Korea Advanced Institute of Science and Technology, Daejeon 34141, Korea}
\affiliation{Center of Complex Systems, KAIST, Daejeon 34141, Korea}

\date{\today}

\begin{abstract}
Supply chains are inherently vulnerable to stochastic shocks due to their sequential dependence on production. Employing the Feld-Barthelemy (FB) framework [Phys. Rev. Lett. {\bf 134} 217401 (2025)], we investigate how capacity volatility and input substitutability determine critical demand that a supply network can withstand. By controlling the variance in production capacity, 
we show that reducing capacity volatility is more effective than increasing average capacity in long supply chains, highlighting the necessity of firm-level stabilization.
In the extended FB model that incorporates substitute suppliers, we find that diversifying suppliers independently increases critical demand, even when average production capacity remains constant. 
Surprisingly, improving network connectivity can also increase critical demand without changing the number of suppliers. This implies that the effect of input substitutability on critical demand is heavily influenced by network topology, even in the absence of stock. Our findings reveal that mitigating capacity volatility and securing structural diversity in supply routes are just as crucial to supply chain resilience as expanding stock capacity.
\end{abstract}

\maketitle

\section{\label{intro} Introduction}

Supply chain management~\cite{mentzer2001defining,stadtler2005supply,bartezzaghi2016journey} aims to reduce costs and improve customer satisfaction by coordinating the flow of materials, products, information and resources from the procurement of raw materials to the final distribution. Efficiency, however, does not necessarily imply stability. Disruptions caused by natural disasters, financial crises, and geopolitical tensions have shown that optimized supply chains can still be vulnerable to systemic breakdowns, motivating extensive studies on supply networks and resilience~\cite{dong2004supply,colon2017economic,caraiani2024production,ramanathan2020sustainable,battiston2007credit,acemoglu2012network,baqaee2018cascading,acemoglu2020endogenous,yang2021robustness,elliott2022networks,elliott2022supply,snyder2016or,gao2016universal}.
 
A key source of supply chain fragility is the sequential dependence among firms, where a local delay or shortage can propagate downstream and amplify across the network~\cite{barrot2016input,carvalho2021supply,buldyrev2010catastrophic}. .
The concept of \textit{timeliness criticality}~\cite{moran2024timeliness} shows that insufficient buffers can drive such systems close to a critical regime, where small local delays trigger delay avalanches. In the context of supply networks, Feld and Barthelemy~\cite{feld2025critical} introduced a stochastic model, in which firms experience random production capacities, and studied how stocks and network structure affect the maximum demand that the network can withstand.

In this paper, we extend the Feld--Barthelemy (FB) model by relaxing two specific assumptions. First, instead of assuming uniformly distributed production capacity with fixed variance, we examine how the volatility of random production capacity affects critical demand. Secondly, in the context of a Leontief (fixed proportions)-type production function that consists of a sum of substitutable inputs, we study how supplier diversification changes supply network performance.

Moreover, we revisit the role of stock capacity and the effect of associated memory in the FB model~\cite{feld2025critical}. By carrying unused inputs over to subsequent periods, stock buffers temporal mismatches between supply and production~\cite{pichler2022forecasting,hallegatte2008adaptive}. However, capacity volatility and input substitutability can offer alternative mechanisms. While low volatility reduces the variance of stochastic fluctuations, input substitution suppresses variability in total available supply through spatial averaging. Finally, we show that these mechanisms can improve critical demand even without stock capacity.

The remainder of this paper is organized as follows: In Sec.~\ref{model}, we describe a stochastic supply network model and discuss how to generalize it with our assumptions. In Sec.~\ref{results}, we analyze the effect of production-capacity volatility on critical demand and examine the role of input substitutability in critical demand. Finally, in Sec.~\ref{conclusion}, we conclude the paper with the summary of our findings and broader implications. Appendices~\ref{appendix-A}--\ref{appendix-E} are provided for detailed mathematical derivations of critical demand in non-substitutable supply chains as well as the asymptotic limits of critical demand in substitutable supply networks. In Appendix~\ref{appendix-F}, we briefly present the supplementary numerical simulations for the possible extension with stock effects and mixed suppliers.

\section{\label{model} Model}

The supply network is represented by a directed acyclic graph (DAG). The DAG begins at a root node ($i=0$) and ends at leaf nodes. Each node $i$ corresponds to a firm that produces its own product. The firm $i$ receives materials from suppliers $\mathcal{S}(i)$, produces its product, and delivers it to customers $\mathcal{C}(i)$. Leaf nodes have no suppliers, {\it i.e.}, $\mathcal{S}(i)=\emptyset$, and can therefore produce regardless of supply. The root node, located at the top of the supply network, delivers its product to external consumers.

At each time step, ordering, production, and delivery proceed in a prescribed sequence. First, external consumers add a constant demand $r$ to the root node. The total demand $D_0(t)$ at the root node includes the unmet demand at the previous time step as follows:
\begin{align}
    D_0(t)=u(t-1)+r.
    \label{eq1}
\end{align}
Here $u(t-1)$ denotes the unmet demand carried over from the previous step, and the initial condition is $u(0)=0$. Given $D_0(t)$, the root node places orders with its suppliers, $l\in\mathcal{S}(0)$, and these orders become the demand of the supplier nodes. 

\begin{figure}[t]
\includegraphics[width=\columnwidth]{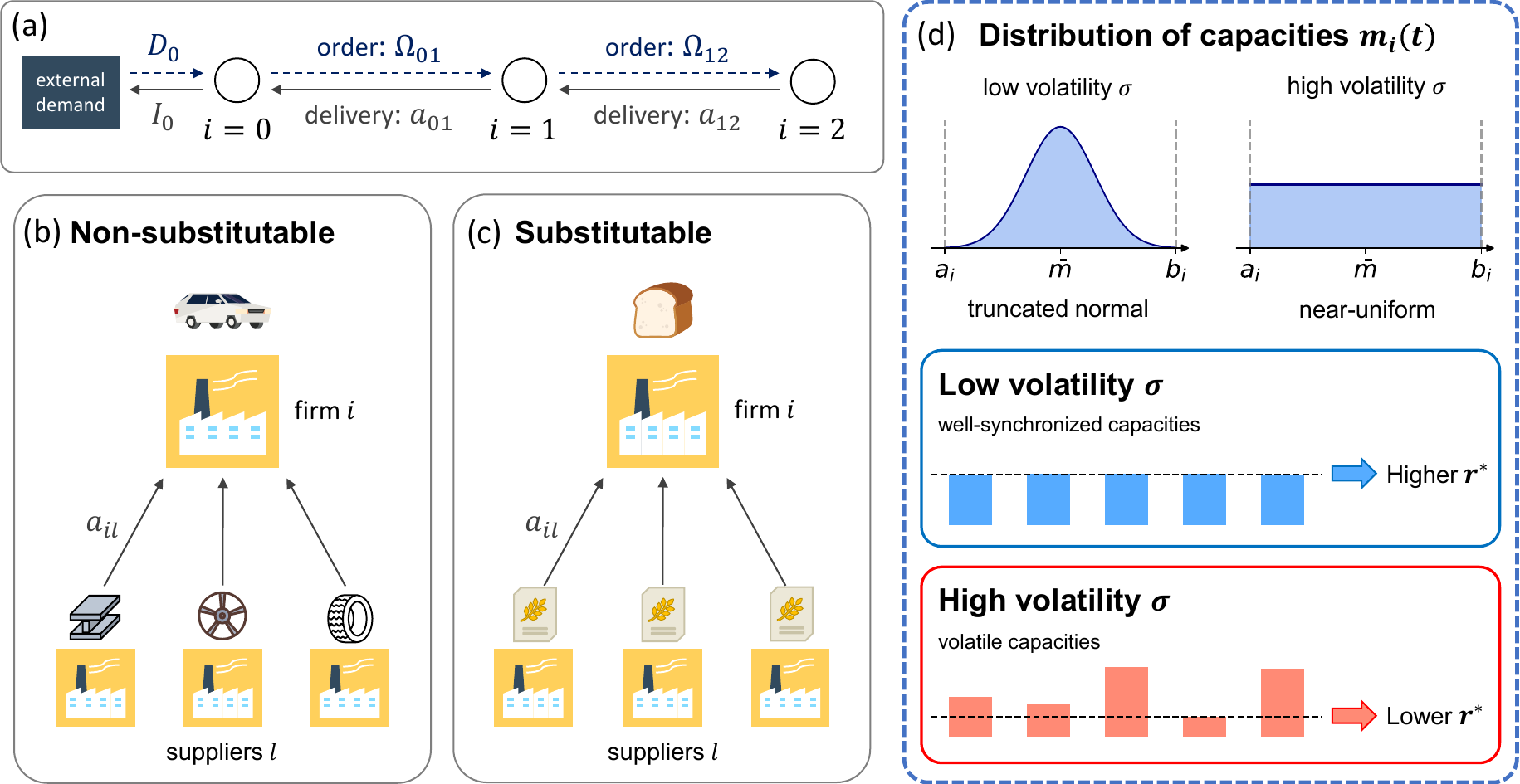}
    \caption{Schematic illustration of order-delivery process and two different inputs: (a) Order propagation (dashed line arrows) and delivery (solid line with arrows) processes in a single chain from root to leaf node. Conceptual comparison of (b) non-substitutable production and (c) substitutable production. Automobile production requires distinct inputs, such as steel, steering wheels, and tires, from different suppliers, whereas bread production can use compatible inputs, such as flour, supplied by multiple firms. (d) Production-capacity distributions with different volatility $\sigma$. As $\sigma$ increases, the truncated normal distribution (left) approaches a near-uniform distribution (right) in the upper panel. Low volatility in the middle panel synchronizes production capacities across firms and weakens bottleneck effects with higher $r^*$, whereas high volatility in the bottom panel makes weak nodes more likely and lowers $r^*$.}
    \label{fig1}
\end{figure}

We here describe only the order-demand propagation process in the non-substitutable supply network. It is noted that the process in the case of a substitute is discussed later in Sec.~\ref{substitute}.

The order $\Omega_{il}(t)$ of node $i$ to its supplier $l\in\mathcal{S}(i)$ and the demand $D_l(t)$ of node $l$ are given by
\begin{align}
    \Omega_{il}(t)
    &=\max\left[0,D_i(t)-k_{il}(t)\right], \label{eq2} \\
    D_l(t)
    &=\sum_{i\in\mathcal{C}(l)}\Omega_{il}(t), \label{eq3}
\end{align}
where $k_{il}(t)$ denotes the stock of product $l$ held by firm $i$. Through these processes, order and demand propagate from a root node toward leaf nodes.

Once demand reaches the leaf nodes, production begins to propagate backward to the root node. The production quantity $I_i(t)$ of node $i$ is determined by the following Leontief-type production function~\cite{pichler2022forecasting}:
\begin{align}
    I_i(t)=\min_{l\in\mathcal{S}(i)}\left[m_i(t),D_i(t),a_{il}(t)+k_{il}(t)\right],
    \label{eq4}
\end{align}
where $m_i(t)$ is the random production capacity of node $i$, and $a_{il}(t)$ is the amount of product delivered from supplier $l$ to firm $i$ at time $t$. For the leaf nodes without suppliers, production is only determined by $m_i(t)$ and $D_i(t)$. The amount of delivery $a_{il}(t)$ is defined as
\begin{align}
    a_{il}(t)=
    \begin{cases}
    0 & \text{if $D_l(t)=0$ or $l\notin\mathcal{S}(i)$}, \\
    \frac{\Omega_{il}(t)}{D_l(t)}I_l(t) & \text{otherwise}.
    \end{cases}
    \label{eq5}
\end{align}
When several customer firms place orders with the same supplier $l$, the output $I_l(t)$ is allocated in proportion to their order quantities $\Omega_{il}(t)$, relative to the total demand $D_l(t)$. The pairs $[I_i(t),a_{il}(t)]$ are then computed sequentially from leaf nodes to the root node [see Fig.~\ref{fig1} (a)].

After production and delivery processes are completed, both stocks and root demand are updated. The stock update rule is 
\begin{align}
    k_{il}(t+1)=
    \begin{cases}
    \min\left[s,a_{il}(t)+k_{il}(t)-I_i(t)\right] & \text{if $l\in\mathcal{S}(i)$}, \\
    0 & \text{otherwise},
    \end{cases}
    \label{eq6}
\end{align}
where $s$ is the stock capacity. Hence, firm $i$ can store the unused quantity of product $l$ after production, up to the capacity limit $s$.

At time $t$, the root node fills external demand $D_0(t)$ through its production $I_0(t)$, so the unmet demand is
\begin{align}
    u(t)=D_0(t)-I_0(t).
    \label{eq7}
\end{align}
In general, the long-term average velocity of unmet demand $v(r)$ as a function of the external demand $r$ is as follows:
\begin{align}
    v(r):=\lim_{t\to\infty}u(t)/t=\max(0,r-r^*),
\end{align}
where $r^*$ is a constant called critical demand and $v(r)$ is singular at $r^*$. By applying Eq.~\eqref{eq7} into Eq.~\eqref{eq1}, the root demand evolves as $D_0(t+1)=D_0(t)-I_0(t)+r$ or equivalently,
\[
\Delta D_0(t)=r-I_0(t).
\]
If the average output of the root node, $\langle I_0\rangle$, is less than the external demand $r$, unmet demand $u(t)$ accumulates on average with the amount of $r-\langle{I_0(t)}\rangle$. Therefore, the maximum external demand that can be sustained by the supply network, referred to as critical demand, is defined as $r^*:=\langle{I_0}\rangle$.

In the FB model~\cite{feld2025critical}, $r^*$ was analytically derived for a tree-like network, including single chains without stock capacity, {\it i.e.}, $s=0$, and it was found that stock capacity increases critical demand. In particular, to show how stock capacity $s$ and the interaction for a given network structure also determine critical demand, uniformly distributed production capacity was assumed, as well as non-substitutable inputs. However, in real supply networks, both the magnitude of firm-level production shocks and the availability of substitutes can vary substantially. Thus, we test the distribution of production capacity and the number of substitute suppliers, and investigate how these two factors affect critical demand, independently of stock capacity.

\section{\label{results} Results}

In this section, we present how production capacity distributions and input substitutability affect critical demand $r^*$ without stock capacity, $s=0$, by analytical and numerical. It is noted that non-zero stock capacity analyses are available in Appendix~\ref{stock_capacity}.

\begin{figure*}[]
\includegraphics[width=\textwidth]{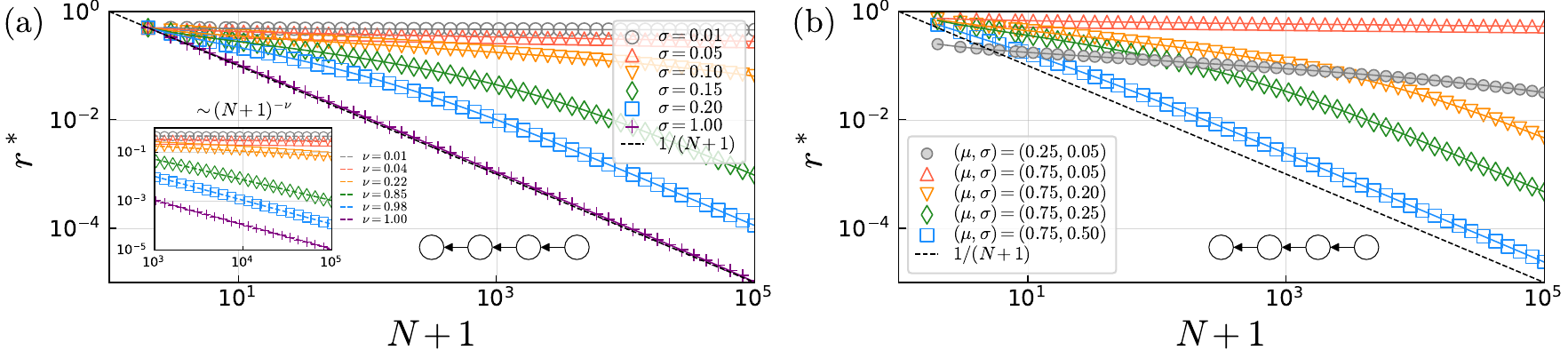}
    \caption{Critical demand $r^*$ against the number of participating firms $N$ in a single chain supply network without stock capacity ($s=0$) and truncated normal-distributed production capacity $m_i(t)\sim{\rm Truncnorm}(\mu,\sigma,0,1)$. Symbols represent numerical simulation results and solid lines represent the analytical expressions of Eqs.~\eqref{eq10} and~\eqref{eq-A3}, evaluated by numerical integration. (a) $\mu=1/2$ for various volatility $\sigma=\{0.01, 0.05, \cdots, 1.00\}$. The inset shows power-law fittings, $r^*\sim(N+1)^{-\nu}$ for sampled range $10^3\leq N+1\leq 10^5$. Dotted lines represent linear regression results on a double-logarithmic scale, and $\nu$ in the right legend shows the estimated exponents. All regression lines satisfy a coefficient of determination $ R^2> 0.99$. (b) $\mu\neq1/2$ for various $\sigma$.}
    \label{fig2}
\end{figure*}

\subsection{\label{capacity} Effect of production capacity distribution}

In the FB model~\cite{feld2025critical}, it is assumed that the production capacity $m_i(t)$ follows an independent and identically distributed ({\it i.i.d.}) uniform random variable on $[0,1]$:
\begin{align}
    m_i(t)\sim{\rm Uniform}(0,1).
    \label{eq9}
\end{align}
In this case, the variance of $m_i(t)$ remains as constant ${\rm Var}(m)=1/12$. However, in real supply networks, the volatility of production capacity can vary substantially depending on the nature of the supply network.

To incorporate this effect, we consider the truncated normal distribution [see Fig.~\ref{fig1} (c)] as follows:
\begin{align}
    f(m;\mu,\sigma,a,b)=\frac{1}{\sigma}\frac{\varphi(\frac{m-\mu}{\sigma})}{\Phi(\frac{b-\mu}{\sigma})-\Phi(\frac{a-\mu}{\sigma})},
    \label{eq10}
\end{align}
where $\varphi$ and $\Phi$ are the probability density function (PDF) and the cumulative distribution function (CDF) of the normal distribution with $\mu$ and $\sigma$, respectively, $\mathcal{N}(\mu, \sigma)$. $a$ and $b$ denote the lower and upper bounds of the support, respectively.
For convenience, we denote $m_i(t)\sim{\rm Truncnorm}(\mu,\sigma,a,b)$ where capacity random variable $m_i(t)$ follows the PDF $f(m;\mu,\sigma,a,b)$ in Eq.~\eqref{eq10}.

Throughout our study, we normalize the upper bounds of external demand and production capacity of the root node to 1. Therefore, the support set for production capacity is fixed as $m_i(t)\in[0,1]$. For the truncated normal distribution, it corresponds to $[a,b]=[0,1]$. It is noted that $f(m;\mu,\sigma,0,1)$ converges to the standard uniform distribution as $\sigma\to\infty$.  

For a single chain without stock capacity, {\it i.e.}, $s=0$, stock is always $k_{i,i+1}(t)=0$ and all nodes have the same demand as the root node, $D_i(t)=D_0(t)$. For chain length $N$, applying the delivery rule $a_{i,i+1}(t)$, recursively, to Eq.~\eqref{eq4} gives $I_i(t)=\min\left[m_i(t),\dots,m_N(t),D_0(t)\right]$. Let $L=N-i$ be the distance from node $i$ to leaf nodes. We define $M_L:=\min\left[m_i,\dots,m_N\right]$ as the minimum production capacity over this segment. By the associativity, we get $I_i(t)=\min\left[M_L(t),D_0(t)\right]$. When demand does not form the bottleneck, $D_0(t)\geq M_L(t)$, the production $I_i(t)$ is simply equivalent to the random variable $M_L$.

Because the production capacity $m_i(t)$ is {\it i.i.d.}, the CDF of $M_L$ can be obtained from that of individual capacities. The derivative yields the corresponding PDF. The average production of node $i$ then follows from this PDF. If $m_i(t)\sim{\rm Uniform}(0,1)$, $\langle I_i\rangle$ is given by the beta integral, resulting in the closed form $\langle I_i\rangle=1/(1+N-i)$. Since critical demand is the average output of the root node, we obtain $r^*=\langle I_0\rangle=1/(N+1)$.

Similarly, for $m_i(t)\sim{\rm Truncnorm}(\mu,\sigma,a,b)$, critical demand $r^*(N;\mu,\sigma,a,b)$ can be derived from the CDF of the minimum-capacity variable [see Appendix~\ref{appendix-A}]. For example, critical demand for $(\mu,a,b)=(1/2,0,1)$ is given by
\begin{widetext}
\begin{align}
     r^*(N,\sigma)=\frac{N}{\left[2{\rm erf}\left(\frac{1}{2\sqrt{2}\sigma}\right)\right]^N}\sqrt{\frac{2}{\pi\sigma^2}}
     \int_{0}^{1}me^{-\frac{(2m-1)^2}{8\sigma^2}}\left[{\rm erf}\left(\frac{1}{2\sqrt{2}\sigma}\right)-{\rm erf}\left(\frac{2m-1}{2\sqrt{2}\sigma}\right)\right]^{N-1}{\rm d}m.
    \label{eq11}
\end{align}
\end{widetext}

Preferentially, we analyze critical demand $r^*$ as capacity volatility $\sigma$ and the average production capacity of firms vary. Fixing $\mu=1/2$, the average production capacity is always $\mathbb{E}[m]=1/2$, regardless of $\sigma$, and is the same as that of the standard uniform distribution in Eq.~\eqref{eq9}.

Figure~\ref{fig2} shows that the theoretical expressions in Eq.~\eqref{eq11} match the numerical simulation results for the various values of $(N,\sigma)$. It shows that if the average production capacity is fixed, the higher volatility always provides the lower critical demand. For sufficiently large volatility, such as $\sigma=1$, critical demand becomes close to $r^*\approx1/(N+1)$ as expected from the convergence of the truncated normal distribution to the standard uniform distribution in the limit $\sigma\to\infty$.

We further consider the capacity distribution as $m_i(t)\sim{\rm Truncnorm}(\mu,\sigma,a=0,b=1)$  for $\mu\neq 1/2$. In this case, the average production capacity is 
\begin{align}
    \mathbb{E}[m]=\mu+\sigma\frac{\varphi(\frac{a-\mu}{\sigma})-\varphi(\frac{b-\mu}{\sigma})}{\Phi(\frac{b-\mu}{\sigma})-\Phi(\frac{a-\mu}{\sigma})}.
    \label{eq12}
\end{align}
Although $\mathbb{E}[m]\neq\mu$, the larger $\mu$ guarantees the larger $\mathbb{E}[m]$ for a given $\sigma$. We compare the case with the smaller average production capacity ($\mu=0.25 < 1/2$) to the case with the larger one ($\mu=0.75 > 1/2$) in Fig.~\ref{fig2}(b), which shows the crossover points between supply chains with the larger average production capacity but the higher volatility and those with the smaller average production capacity but the lower volatility. This indicates a trade-off between production performance (higher $\mathbb{E}[m]$) and stability (lower $\sigma$) that depends on the supply chain length $N$. For large $N$, a supply chain with the lower average capacity but the smaller volatility can sustain the larger critical demand.

In addition, we describe the scaling behavior of critical demand for large $N$ under truncated normal-distributed capacities. The inset in the Fig.~\ref{fig2}(a) shows the power-law fittings, $r^* \sim (N+1)^{-\nu}$, applied to data samples obtained within the range $10^{3} \leq N \leq 10^{5}$. It seems that, for small $N$, critical demand $r^*$ decreases with a slope less than $\nu$. However, for supply chains with higher volatility $\sigma$, this slope shifts more sharply, ultimately appearing to converge to a power-law form with $\nu=1$.

For very large $N$, the asymptotic behavior near the lower bound of the production capacity distribution $\lim_{m\to0^+}f(m)$ dominates critical demand $r^*$~\cite{fisher1928limiting,manzato2012fracture,majumdar2020extreme}. Actually, for the case of truncated normal-distributed capacities, critical demand $r^*$ for $N\to\infty$ scales as
\begin{align}
    r^*\sim N^{-1}\quad\text{for $N\to\infty$}.
\end{align}
This relationship is independent of the parameters $(\mu,\sigma)$ (see Appendix~\ref{appendix-B} for detailed mathematical derivations). Furthermore, the approximate scale at which system enters this power-law regimes, such that
\begin{align}
    N_{\rm power}\sim\frac{\mu}{\sigma^2f(0)}.
\end{align}
For $\mu=0.5$, $N_{\rm power}\sim3.4\times 10^6$ at $\sigma=0.1$; $1.41\times 10^2$ at $\sigma=0.2$ (see Appendix~\ref{appendix-B} for more derivations). It is noted that the following two limits are not interchangeable.
\begin{align}
    \lim_{N\to\infty}\lim_{\sigma\to0^+}r^*(N,\sigma)=\mu, \\
    \lim_{\sigma\to0^+}\lim_{N\to\infty}r^*(N,\sigma)=0.
\end{align}
As a result, this highlights the distinction between quasi-deterministic behaviors for the extremely low volatility $\sigma$ and extreme-value-dominated behaviors for the extremely large supply chain length $N$.

The power-law behavior of $r^*$ for the large $N$ limit, $r^*\sim(N+1)^{-1}$, does not hold for all distributions. Finally, we consider such a case, in which the production capacity follows a beta distribution:
\begin{align}
    m_i(t)\sim{\rm Beta}(\alpha,\beta),
\end{align}
where the support is $m\in[0,1]$, $\alpha$ and $\beta$ are positive constants, and the average and variance are $\mathbb{E}[m] = \frac{\alpha}{\alpha+\beta}$ and $\text{Var}(m) = \frac{\alpha\beta} {(\alpha+\beta)^2(\alpha+\beta+1)}$, respectively. It is noted that ${\rm Beta}(1,1)$ is identical to the standard uniform distribution. In particular, we focus on critical demand for the case that $m_i(t) \sim \text{Beta}(\alpha, \beta=\alpha)$, $\mathbb{E}[m] = 1/2$, but $\text{Var}(m)$ decreases as $\alpha$ increases.

\begin{figure}[]
\includegraphics[width=\columnwidth]{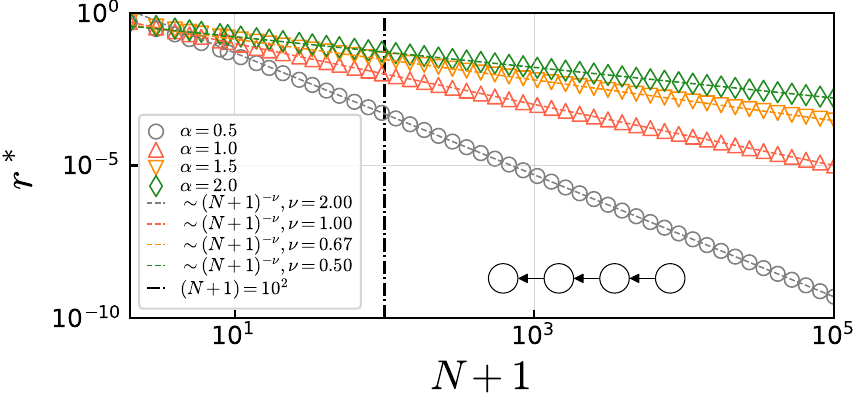}
    \caption{Critical demand $r^*$ against the number of firms $N$ in a single-chain supply network for the case of beta-distributed capacities, $m_i(t)\sim{\rm Beta}(\alpha,\alpha)$. Symbols represent numerical simulation results, and dotted lines represent linear regression results from samples with $(N+1)\geq10^2$. All regression lines satisfy the coefficient of determination of $R^2>0.999$.}
    \label{fig3}
\end{figure}

In the case of beta-distributed capacities, the asymptotic scaling behavior of $r^*$ for $N\to\infty$ becomes
\begin{align}
    r^*\sim N^{-1/\alpha}\quad\text{for $N\to\infty$}.
    \label{eq18}
\end{align}
Figure~\ref{fig3} shows the asymptotic scaling behavior of critical demand $r^*$ for the case of beta-distributed capacities, where we observe that power-law decay can be even faster ($\nu >1$) or slower ($\nu <1$) than $(N+1)^{-1}$, unlike the case of a truncated normal distribution [see Appendix~\ref{appendix-C}].

In this subsection, we present how the production capacity distribution affects the scaling behavior of critical demand in supply network without substitutable inputs and stock capacity ($s=0$). Our results are obtained in a single chain by numerical simulations, Eqs.~\eqref{eq11} and~\eqref{eq18}, but they still hold for a tree where $|\mathcal{C}(i)|=1$ for a non-root node $i$. Under these circumstances, critical demand $r^*$ depends only on the number of firms $N$.

\subsection{\label{substitute} Effect of input substitutability}

The production function in Eq.~\eqref{eq4} describes the case that all suppliers $l$ of firm $i$ provide non-substitutable inputs. However, in real supply networks, multiple firms may produce compatible inputs~\cite{elliott2022networks,tomlin2006value}. 

In order to examine the role of substitute suppliers, we consider that all suppliers $l\in\mathcal{S}(i)$ of firm $i$ provide substitutable inputs. The production function is then given by
\begin{align}
    I_i(t)=\min\left[m_i(t),D_i(t),\sum_{l\in\mathcal{S}(i)}a_{il}(t)+k_i(t)\right].
    \label{eq19}
\end{align}
Because inputs are substitutable, stocks are no longer distinguished by supplier and indexed only by firm $i$. 

The stock update rule becomes
\begin{align}
    k_i(t+1)=\min\left[s,\sum_{l\in\mathcal{S}(i)}a_{il}(t)+k_i(t)-I_i(t)\right].
    \label{eq20}
\end{align}
The demand $D_i(t)$ and the delivery quantity $a_{il}(t)$ are defined in Eqs.~\eqref{eq3} and~\eqref{eq5}, respectively. 

However, the order quantity $\Omega_{il}(t)$ differs from the non-substitutable case. The firm $i$ divides the required input quantity $D_i(t)-k_i(t)$ among its suppliers, so that $\sum_{l\in\mathcal{S}(i)}\Omega_{il}(t)=D_i(t)-k_i(t)$. If the order is divided equally among its suppliers, the order quantity becomes 
\begin{align}
    \Omega_{il}(t)=\max\left(0,\frac{D_i(t)-k_i(t)}{|\mathcal{S}(i)|}\right),
    \label{eq21}
\end{align}
where $|\mathcal{S}(i)|$ is the number of suppliers of firm $i$.

Suppose firm $i$ receives inputs from leaf suppliers, 
$l=\{j \mid j \in \mathcal{S}(i) \text{ and } \mathcal{S}(j) = \emptyset\}$.
If demand is not a bottleneck, its output is $I_i(t)=\min[m_i,\sum_{l\in\mathcal{S}(i)} m_l]$. 
The aggregate production capacity, $Y_i=\sum_{l\in\mathcal{S}(i)} m_l$, can be treated as a new random variable. The main difference from the non-substitutable case is that the random variables compared in the minimum function no longer follow identical distributions.

If individual production capacities $m_i$ and $m_l$ follow the same distribution, then $m_i$ and $Y_i$ have different supports. Because of the additive structure, the average aggregate capacity, $\mathbb{E}[Y_i]$, increases by a factor of $|\mathcal{S}(i)|$, so that firm $i$'s output depends mainly on $m_i$. This represents a situation in which the suppliers' aggregate production capacity exceeds that of their customer.

For a supply network, the depth of node $i$, denoted by $d_i$, is its shortest distance from the root. The set of nodes at the same depth $d$ is called layer $\mathcal{L}(d)$. The number of possible depth values, including the root node at $d=0$, is height $h$ of the supply network.
In order to match the aggregate production capacity level for each layer, we restrict the support as
\begin{align}
    \sum_{i\in\mathcal{L}(d)} m_i\in [0,1].
    \label{eq22}
\end{align}
This implies that production scale is normalized across all layers. 
Assuming that all individual production capacities $m_i(t)$ are independent and uniformly distributed,
\begin{align}
    m_i(t)\sim{\rm Uniform}(0,b_i),
    \label{eq23}
\end{align}
where $b_i$ is the upper bound of the support for node $i$ and $b_0=1$. Alternatively, $m_i(t)=b_iU_i$, where $U_i$ is {\it i.i.d.} standard uniform random variable. 

In this subsection, the upper bound is
\begin{align}
    b_i=1/|\mathcal{L}(d_i)|.
    \label{eq24}
\end{align}
Accordingly, the aggregate capacity for each layer always satisfies Eq.~\eqref{eq22} and its average is fixed at $1/2$. 

If the supply network is a regular tree, the upper bound of the capacity for firm $i$ and its supplier $l\in\mathcal{S}(i)$ have the following relationship:
\begin{align}
    b_l=b_i/|\mathcal{S}(i)|\quad\text{where } l\in\mathcal{S}(i).
\end{align}
This rule can be applied sequentially from a root node to leaf nodes.
\begin{figure}[]
\includegraphics[width=\columnwidth]{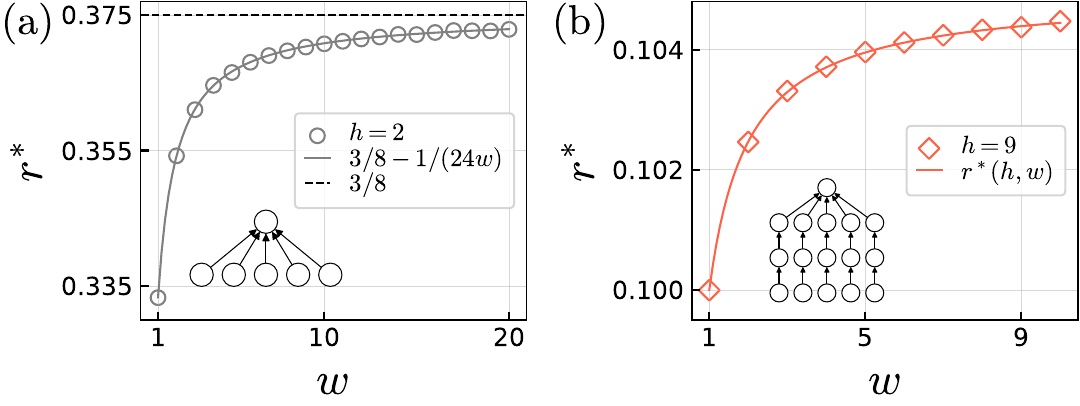}
    \caption{Critical demand $r^*$ against width $w$ in a multiple-chain supply network with fixed heights (a) $h=2$ and (b) $h=9$. Symbols and solid lines represent numerical simulation results and analytical results by Eqs.~\eqref{eq28}-\eqref{eq29}, respectively. Insets show schematic illustrations of the network structure, ($h=2, w=5$) and ($h=4, w=5$), respectively.}
    \label{fig4}
\end{figure}
\begin{figure}[]
\includegraphics[width=\columnwidth]{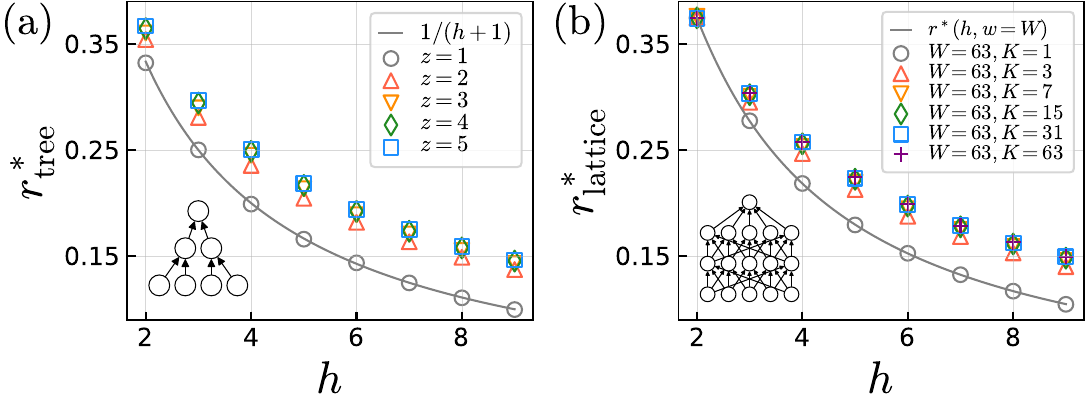}
    \caption{Critical demand $r^*$ against height $h$ in regular-tree structure (a) for various branching factors $z$ and (b) for various width $W$ and connectivity $K$. Symbols and solid lines represent numerical simulation results and analytical results by Eq.~\eqref{eq29}, respectively. Insets show schematic illustrations for the network structure of ($h=3, z=2$) and that of ($h=4, W=5, K=3$), respectively.}
    \label{fig5}
\end{figure}
%
In multiple-chain structures, several single chains are connected to the root node. Let $h$ denote height and $w$ denote width, equivalent to the number of suppliers per layer. For $h=2$, if demand is not a bottleneck, then the output of the root node is as follows:
\begin{align}
    I_0(t)
    &=\min[m_0(t),D_0(t),\sum_{l\in\mathcal{S}(0)}a_{0l}(t)] \nonumber \\
    &=\min[m_0(t),\sum_{l\in\mathcal{S}(0)}m_l(t)].
\end{align}
Applying Eqs.~\eqref{eq21} and~\eqref{eq23} to this network, individual and aggregate production capacities are given by
\begin{align}
    \begin{cases}
    m_0(t) &\sim {\rm Uniform}(0,1), \\
    m_l(t) &\sim {\rm Uniform}(0,1/w), \\
    w\sum_{l=1}^{w}m_l(t) &\sim \text{Irwin--Hall}(w).
    \end{cases}
\end{align}
The random variable $m_0$ and the scaled random variable $wm_l$ follow the standard uniform distribution. Thus, the scaled aggregate capacity is the sum of $w$ independent standard uniform random variables,
$$\omega Y_0=w\sum_{l=1}^{\omega}m_l,$$ 
which follows an Irwin--Hall distribution.
The average output of the root node is $\mathbb{E}[I_0]=\mathbb{E}[\min(m_0,Y_0)]$. 
Since both $m_0$ and $Y_0$ have the same support, the conditional average at $Y_0=y$ is as follws: 
$$\mathbb{E}[\min(m_0,Y_0)|Y_0=y]=\int_{0}^{y}m\text{d}m+(1-y)y=y-y^2/2.$$ 
Hence $\mathbb{E}[\min(m_0,Y_0)]=\mathbb{E}[Y_0]-\mathbb{E}[Y_0^2]/2$. 
Employing the moments of the Irwin--Hall distribution, we obtain
\begin{align}
    r^*(w)=\frac{3}{8}-\frac{1}{24w}.
    \label{eq28}
\end{align}

In contrast to the non-substitutable case that critical demand $r^*$ decreases as the number of required suppliers $N$ increases, all inputs to the root node are substitutable, $r^*$, instead increases with width $w$ and approaches $3/8$ for $w\to\infty$ [see Fig.~\ref{fig4}(a)].
%
This enhancement can be explained by the consequence of the spatial averaging of stochastic production capacities. 
Let $Y^{(w)}=\sum_{l\in\mathcal{S}(i)}m_l$ denote suppliers’ aggregate production capacity. 
While the average remains $\mathbb{E}[Y^{(w)}]=1/2$, the variance decreases to ${\rm Var}(Y^{(w)})=1/(12w)$. Thus, Eq.~\eqref{eq28} can be rewritten as $r^*=3/8-{\rm Var}(Y^{(w)})/2$. As $w$ increases, $r^*$ gets larger by reducing fluctuation-induced production losses without increasing the average aggregate capacity (see Appendix~\ref{appendix-D} for detailed derivations).

For general $h$, using the moments of the sum of independent beta-distributed random variables, critical demand is given by:
\begin{align}
    r^*(h,w)=\frac{1}{h}\left(1-\frac{1}{2h}\right)-\frac{(h-1)}{2h^2(h+1)w}.
    \label{eq29}
\end{align}
Increasing $h$ induces more inter-layer (sequential) dependencies, so that $r^*$ gets smaller. In contrast, increasing $w$ induces more substitute chains, so that $r^*$ gets larger [see Fig.~\ref{fig4}(b)].

In Fig.~\ref{fig5}, we present $r^*$ in a regular tree network, where every node has $z$ suppliers and $b_i = 1/z^{d_i}$ for all nodes. 
%
Figure~\ref{fig5}(a) shows the simulated critical demand $r^*$ for various values of $(h,z)$. To examine the limits of the input substitution effect, we choose the following infinite-branching limit,
\begin{align}
    R_h := \lim_{z\to\infty} r^*(h, z).
\end{align}
In a regular tree, every supplier node has exactly one consumer node, such that $|\mathcal{C}(l)|=1$. 

Consequently, $a_{il}(t) = I_l(t)$. When demand is not a bottleneck, the production of node $i$ is given by
\begin{align}
    I_i^{(d)}=\min\left[m_i^{(d)},\sum_{l\in\mathcal{S}(i)} I_l^{(d+1)}\right].
\end{align}
Here $I_i^{(d)}$ and $m_i^{(d)}\sim{\rm Uniform}(0, 1/z^d)$ denote the production and production capacity for node $i$ in layer $d$, respectively. This yields a nonlinear recurrence relation for $r^*$ (see Appendix~\ref{appendix-D} for more detailed mathematical derivations).
\begin{align}
    R_{h+1}=R_h-\frac{R_h^2}{2},\quad R_1=\frac{1}{2}.
    \label{eq32}
\end{align}

\begin{figure}[]
\includegraphics[width=\columnwidth]{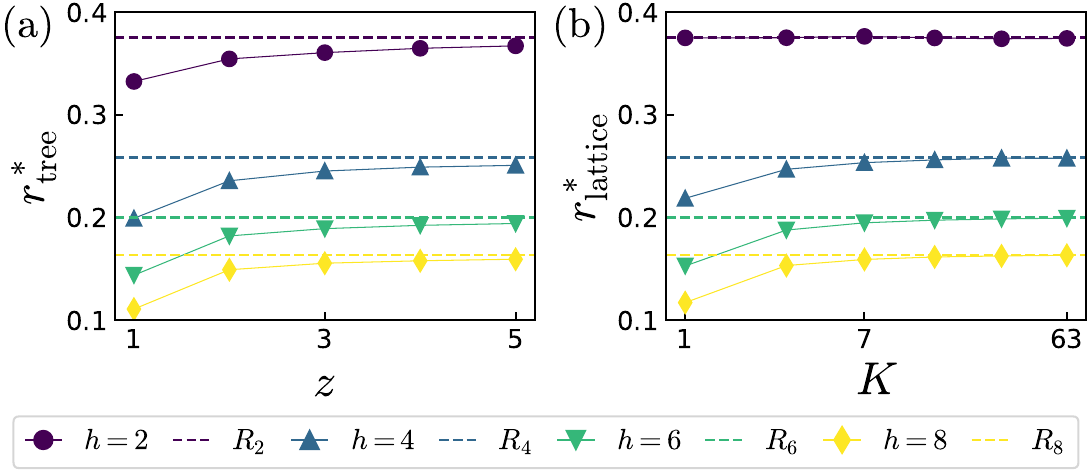}
    \caption{Critical demand $r^*$ for various $h=\{2,4,6,8\}$ against (a) branching factor $z$ in a regular tree and (b) connectivity $K$ in a regular lattice, respectively. Symbols in (a) and (b) correspond to numerical simulation results in Fig.~\ref{fig5}~(a) and (b). Dotted lines show the analytical results of $R_h$ by Eq.~\eqref{eq32}.}
    \label{fig6}
\end{figure}

Using this recurrence, we determine $r^*$ for any height $h$ in the infinite-branching limit $z\to\infty$. 
Figure~\ref{fig6}(a) shows that $r^*$ converges to its theoretical limit as $z$ increases.

When we consider a directed regular lattice network, in which every layer except the root layer contains $W$ nodes, each with $K$ suppliers and consumers. In this case, $b_i = 1/W$ for all nodes except the root.

Unlike a regular tree, where increasing $z$ also increases the number of participating firms, $N=z^{h-1}$, this setting allows us to vary the network connectivity $K$ while keeping the number of nodes $N=1+(h-1)W$ fixed. Thus, we can isolate a topological effect on $r^*$ independently of $h$ and $N$. Figure~\ref{fig5}(b) shows the simulated critical demand $r^*$ for various values of $(h,K)$. The results demonstrate that increasing $K$ alone can improve $r^*$, even without adding substitute suppliers to the network.

Interestingly, Fig.~\ref{fig6}(b) shows that in the fully connected, infinite-width limit ($K=W$ and $W\to\infty$), $r^*$ converges to $R_h$, the infinite-branching limit for a regular tree, as given by Eq.~\eqref{eq32}, despite the different network topologies (see the Appendix~\ref{appendix-E} for detailed derivations).

\begin{figure}[t]
\includegraphics[width=\columnwidth]{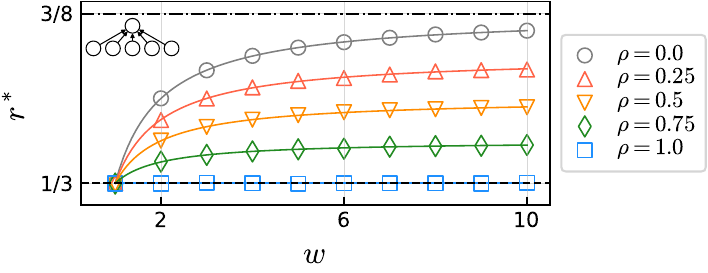}
    \caption{Critical demand $r^*$ against $w$ in a multiple-chain structure at height $h=2$ for various capacity correlation $\rho$. Symbols and solid lines represent numerical simulation results and analytical result by Eq.~\eqref{eq34}, respectively. The inset shows the schematic illustration of the network structure at $w=5$.}
    \label{fig7}
\end{figure}

Finally, we extend the analysis of correlations among the production capacities of substitute suppliers, where we employ a multiple-chain supply network with $h=2$ and assume that the Pearson correlation coefficient between the capacities of leaf suppliers is constant:
\begin{align}
    {\rm Corr}(m_k(t),m_l(t))=\rho\quad\text{for $k\neq l$},
\end{align}
where $0\leq\rho\leq1$ and $k,l\in\mathcal{S}(0)$. 
The capacities of the root and supplier nodes remain independent, and each capacity follows a uniform distribution with an upper bound of $b_i=1/w^{d_i}$. Hence, $r^*$ is
\begin{align}
    r^*(w,\rho)=\frac{3}{8}-\frac{1+(w-1)\rho}{24w}.
    \label{eq34}
\end{align}
Even when width $w$ is fixed, $r^*$ decreases as the capacity correlation $\rho$ increases [see Fig.~\ref{fig7}]. 
This suggests that correlations among production capacities reduce the dispersion of stochastic shocks. When capacities are perfectly correlated, $r^*(w,\rho=1)=1/3$, the supply network is equivalent to the single chain with height $h=2$ and the substitution effect disappears entirely.

\section{\label{conclusion} Conclusion}
To sum up, we presented a stochastic supply network model and investigated the role of production capacity volatility and input substitutability in critical demand $r^*$:
First, we found that higher volatility $\sigma$ in production capacity lowers critical demand $r^*$, which becomes more pronounced as the number of firms $N$ increases. In supply chains where a disruption at one node can create a system-wide bottleneck~\cite{manzato2012fracture,majumdar2020extreme}, our finding suggests that stabilizing firms’ production capacities at any given time is more important than simply increasing their average capacities. We also analyzed $r^*$ in the limit $N\to\infty$, and showed that the asymptotic scaling behavior of $r^*$ depends not only on the variance but also on how the distribution behaves near the lower bound. Specifically, $r^* \sim N^{-1}$ for uniform and truncated normal distribution, and $r^* \sim N^{-1/\alpha}$ for a beta distribution.
Second, we showed that input substitutability raises critical demand, even when the average production capacity in each layer remains identical. The core mechanism is that the stochastic production shocks are dispersed across independent suppliers, reducing the impact of individual fluctuations. Correlations among production capacities of suppliers, by contrast, weaken this mechanism and decrease critical demand. Thus, supplier diversification (spatial averaging) can help spread risk, but its effectiveness depends on whether suppliers' productions are truly independent~\cite{mizgier2015disentangling}. Moreover, when inputs are substitutable, critical demand depends not only on the number of substitute suppliers but also on network topology, including connectivity~\cite{radicchi2017redundant}.

Real-world supply chain disruptions illustrate the importance of alternative suppliers and routes. The disruption of Japan’s automotive industry after the 2011 earthquake~\cite{matsuo2015implications,inoue2019firm,carvalho2021supply} and global maritime bottlenecks caused by the 2021 Suez Canal incident~\cite{tran2025costs} show how localized problems can escalate into system-wide disruptions when substitution options are limited, or supply routes are concentrated in particular segments (see Appendix~\ref{appendix-F} for complementary numerical simulations of extra cases). 

Our study offers a minimal theoretical mechanism to explain real phenomena. The implication of our results is especially relevant to just-in-time (JIT) production systems, in which firms avoid large stocks to reduce costs~\cite{tomlin2006value,graves2000optimizing}. In such systems, coordinating delivery schedules reliably and securing alternative suppliers are important ways to mitigate stochastic production shocks.

Several avenues for future research remain: 
First, firms that share infrastructure, such as power grids, telecommunications systems, and transportation networks, may have correlated production capacities. We considered the simple case in which all substitute suppliers share an identical correlation coefficient, but correlations in reality may be distributed more complexly in practice and may be auto-correlated. 
Second, firms might allocate orders adaptively, based on supplier reliability or expected production capacity~\cite{battiston2007credit}, rather than distributing them equally. They might also adjust their target production~\cite{martin2026resilient} accordingly. 
Finally, incorporating financial constraints, pricing, and competition among substitute suppliers could extend our model study to address firms’ strategic behavior and profitability, in addition to physical production constraints.

\begin{acknowledgments} 
This research was supported by Basic Science Research Program through the National Research Foundation of Korea (NRF) (KR) [NRF-RS-2026-25489888~(J.H., M.H.) and NRF-RS-2025-00514776~(J.H., H.J.)].
\end{acknowledgments} 

\begin{appendix}
\setcounter{figure}{0}
\setcounter{table}{0}
\renewcommand{\thefigure}{A\arabic{figure}}
\renewcommand{\thetable}{A\arabic{table}}

\section{\label{appendix-A} Derivation of critical demand\\ for a non-substitutable supply network\\ without stock capacity in a tree}

For a non-substitutable supply network with no stock capacity ($s=0$) in a tree, critical demand $r^*$ is the average output of the root node under sufficient demand. Applying $D_0\geq m_i$ to Eq.~\eqref{eq4} gives
\begin{align}
    r^*:=\langle I_0\rangle=\mathbb{E}[\min(m_0,m_1,\dots,m_{N-1})],
    \label{eq-A1}
\end{align}
where $N$ is the number of nodes in the chain. Let $m_i$ be i.i.d. random variables and $f(m)$ and $F(m)$ be the probability density function (PDF) and cumulative distribution function (CDF). For $I_0=\min(m_0,\dots,m_{N-1})$, we have
\begin{align*}
    \mathbb{P}(I_0>m)=\mathbb{P}(m_0>m,\dots,m_{N-1}>m)=[1-F(m)]^N.
\end{align*}
Thus, the CDF and PDF of $I_0$ are
\begin{align*}
    F_{I_0}(m)&=1-[1-F(m)]^N, \\
    f_{I_0}(m)&=Nf(m)[1-F(m)]^{N-1}.
\end{align*}
The average root output is therefore
\begin{align}
    \langle I_0\rangle=N\int_a^b mf(m)[1-F(m)]^{N-1}dm,
    \label{eq-A2}
\end{align}
where $[a,b]$ is the support of $m_i$.
For a truncated normal distribution,
\begin{align*}
    f(m)
    &=\frac{1}{\sigma}\frac{\varphi\left(\frac{m-\mu}{\sigma}\right)}{\Phi(\frac{b-\mu}{\sigma})-\Phi(\frac{a-\mu}{\sigma})}, \\
    F(m)
    &=\frac{\Phi(\frac{m-\mu}{\sigma})-\Phi(\frac{a-\mu}{\sigma})}{\Phi(\frac{b-\mu}{\sigma})-\Phi(\frac{a-\mu}{\sigma})}.
\end{align*}
where $\varphi$ and $\Phi$ are the PDF and CDF of the normal distribution. Substituting these expressions into Eq.~\eqref{eq-A2} yields
\begin{align}
    r^*(N;\mu,\sigma,a,b)=\frac{N}{\left[{\rm erf}\left(\frac{b-\mu}{\sqrt{2}\sigma}\right)-{\rm erf}\left(\frac{a-\mu}{\sqrt{2}\sigma}\right)\right]^N}\sqrt{\frac{2}{\pi\sigma^2}} \nonumber \\
    \int_a^b me^{-\frac{(m-\mu)^2}{2\sigma^2}}\left[{\rm erf}\left(\frac{b-\mu}{\sqrt{2}\sigma}\right)-{\rm erf}\left(\frac{m-\mu}{\sqrt{2}\sigma}\right)\right]^{N-1}dm.
    \label{eq-A3}
\end{align}
The solid lines in Fig.~\ref{fig2}(b) are obtained from numerically evaluating Eq.~\eqref{eq-A3}.

\section{\label{appendix-B} Asymptotic scaling behavior\\ for the case of truncated normal distribution}

For a non-substitutable supply network with no stock capacity ($s=0$), critical demand is given by
\[
r^*=\mathbb{E}[\min(m_0,\dots,m_{N-1})],
\]
as in Eq.~\eqref{eq-A1}. As $N$ increases, the minimum approaches the lower bound of the support. In the limit $N\to\infty$, the behavior of the PDF $f(m)$ near this bound—$\lim_{m\to a^+}f(m)$—plays a crucial role.

Suppose the production capacity has support $[a,b]=[0,1]$. Since the CDF satisfies $F(0)=0$, critical demand can be written as:
\begin{align}
    r^*(N)=\int_{0}^{1}[1-F(m)]^Ndm.
    \label{eq-B1}
\end{align}
Integration by parts shows that this expression is equivalent to Eq.~\eqref{eq-A2}. Assume that $f(m)$ is smooth near $m=0$ and that $0<f(0)<\infty$. Then, for sufficiently small $m$:
\begin{align}
    F(m)=f(0)m+O(m^2).
\end{align}
Note that for a standard uniform distribution, $F(m)=f(0)m$. The integral in Eq.~\eqref{eq-B1} can then be approximated as follows:
\begin{align}
    \int_{0}^{1}[1-F(m)]^Ndm\approx\int_{0}^{1}e^{-Nf(0)m}dm.
\end{align}
Consequently, critical demand behaves as:
\begin{align}
    r^*(N)\sim\frac{1}{Nf(0)}\quad\text{for $N\to\infty$}.
\end{align}
Next, we estimate the scale of $N$ at which this asymptotic power-law behavior begins to emerge. Expanding the truncated normal distribution around $m=0$ gives $f(m)=f(0)[1+(\mu/\sigma^2)m+O(m^2)]$. Thus,
\begin{align}
    f(m)\approx f(0)\left(1+\frac{\mu}{\sigma^2}m\right).
\end{align}
Near $m=0$, the characteristic length scale over which the probability density varies is $\ell=\sigma^2/\mu$. For $f(m)\approx f(0)$ to hold, the typical minimum must be much smaller than this scale: $M_N\sim[Nf(0)]^{-1}\ll\sigma^2/\mu$. Therefore, the approximate scale at which critical demand begins to follow a power-law is:
\begin{align}
    N_{\rm power}\sim\frac{\mu}{\sigma^2f(0)}.
    \label{eq-B6}
\end{align}
This estimate applies when $\mu>0$ and is particularly effective when $\sigma\ll\mu$.

\section{\label{appendix-C} Asymptotic scaling behavior\\ for the case of beta distribution}

For a random variable $X \sim \text{Beta}(\alpha, \beta)$, the probability density function $f(m;\alpha,\beta)$ is:
\begin{align}
    f(m;\alpha,\beta)=\frac{m^{\alpha-1}(1-m)^{\beta-1}}{{\rm B}(\alpha,\beta)}.
\end{align}
where $m \in [0,1]$, $\alpha$ and $\beta$ are positive constants, ${\rm B}(\alpha,\beta) = \Gamma(\alpha)\Gamma(\beta)/\Gamma(\alpha+\beta)$, and $\Gamma$ is the gamma function.

Unlike the uniform and truncated normal distributions, whose CDFs are linear near the lower bound, $F(m) \sim Am$, the CDF of the beta distribution behaves as follows near its lower bound:
\begin{align}
    F(m) \sim Am^{\alpha}\quad\text{for $m \to 0^+$}.
\end{align}
Using this behavior to approximate the expression in Eq.~\eqref{eq-B1} for a non-substitutable supply network with no stock capacity gives:
\begin{align}
    r^*(N) \approx \int_{0}^{1} e^{-NAm^\alpha} dm = \frac{1}{\alpha(NA)^{1/\alpha}} \gamma\left(\frac{1}{\alpha}, NA\right).
\end{align}
Here $\gamma$ denotes the lower incomplete gamma function. Since $\lim_{x \to \infty} \gamma(1/\alpha, x) = \Gamma(1/\alpha)$, as $N \to \infty$ we have $r^* \approx \frac{\Gamma(1/\alpha)}{\alpha(NA)^{1/\alpha}}$. Thus, the asymptotic scaling behavior of critical demand is:
\begin{align}
    r^*(N) \sim N^{-1/\alpha} \quad\text{for $N \to \infty$}.
\end{align}

\section{\label{appendix-D} Critical demand in a regular tree\\ for infinite-branching limit}

We consider a substitutable supply network with no stock capacity ($s=0$). Recall the Eqs.~\eqref{eq19},~\eqref{eq5},~\eqref{eq21}, and~\eqref{eq3}, and by applying $k_i(t)=0$ when demand is not a production bottleneck, we obtain the production $I_i(t)$, delivery $a_{il}(t)$, order $\Omega_{il}(t)$, and demand $D_l(t)$ relations between node $i$ and its supplier $l\in\mathcal{S}(i)$ as follows:
\begin{align}
    \begin{cases}
        I_i(t)
        &=\min[m_i(t),\sum_{l\in\mathcal {S}(i)}a_{il}(t)], \\
        a_{il}(t)
        &=\frac{\Omega_{il}(t)}{D_l(t)}I_l(t), \\
        \Omega_{il}(t)
        &=\frac{D_i(t)}{|\mathcal{S}(i)|}, \\
        D_l(t)
        &=\sum_{i\in\mathcal{C}(l)}\Omega_{il}(t).
    \end{cases}
    \label{eq-D1}
\end{align}
In a regular tree, each supplier $l\in\mathcal{S}(i)$ has exactly one customer, so $|\mathcal{C}(l)|=1$. Thus, $D_l(t)=\Omega_{il}(t)$ and $a_{il}(t)=I_l(t)$. Through this, we define
\begin{align}
    I_i^{(d)}:=\min\left[m_i^{(d)},\sum_{l\in\mathcal{S}(i)}I_l^{(d+1)}\right],
    \label{eq-D2}
\end{align}
where $I_i^{(d)}$ and $m_i^{(d)}\sim{\rm Uniform}(0,1/z^d)$ denote the output and production capacity of node $i$ at depth $d$. Each capacity $m_i^{(d)}$ follows Eq.~\eqref{eq23} and~\eqref{eq24}; hence,
\begin{align}
    m_i^{(d)}=z^{-d}U_i\sim{\rm Uniform}(0,1/z^d).
\end{align}
The root is always at depth $d=0$, and the leaves are at depth $d=h-1$. Thus, $I_i^{(0)}=I_0$ and $I_i^{(h-1)}=m_i^{(h-1)}$.

Multiplying both sides of Eq.~\eqref{eq-D2} by $z^d$ gives
\begin{align}
    z^{d}I_i^{(d)}=\min\left[z^dm_i^{(d)},\frac{1}{z}\sum_{l\in\mathcal{S}(i)}z^{d+1}I_l^{(d+1)}\right].
\end{align}
Here $z^d m_i^{(d)} = U_i^{(d)}$, where $U_i^{(d)}$ is the normalized production capacity of node $i$ at depth $d$ and follows the standard uniform distribution. In the infinite-branching limit $z \to \infty$, the law of large numbers gives:
\begin{align}
    \frac{1}{z}\sum_{l\in\mathcal{S}(i)}z^{d+1}I_l^{(d+1)}\xrightarrow{z\to\infty}\mathbb{E}[z^{d+1}I_l^{(d+1)}].
\end{align}
Thus, the average normalized output at depth $d$ satisfies
\begin{align}
    \mathbb{E}[z^{d}I_i^{(d)}]=\mathbb{E}[\min(U_i^{(d)},\mathbb{E}[z^{d+1}I_j^{(d+1)}])].
    \label{eq-D6}
\end{align}
By repeatedly applying this relation to the right-hand side, we obtain the average production of node $i$ at depth $d$ as follows:
\begin{align}
    \mathbb{E}[I_i^{(d)}]=\frac{1}{z^d}\mathbb{E}[ \min(U_i^{(d)}, \mathbb{E}[ \min(U_j^{(d+1)}, \dots)] ) ].
\end{align}
For $d=0$, this equation is equivalent to critical demand $r^*=\mathbb{E}[I_0]$.

For convenience, we define the function
\begin{align}
    g(x):=\mathbb{E}[\min(U,x)],\quad U\sim{\rm Uniform}(0,1),
    \label{eq-D8}
\end{align}
where $0\leq x\leq 1$. This function captures the relationship between average normalized outputs and normalized aggregate inputs between the successive customers and suppliers in layers $d$ and $d+1$ [see Eq.~\eqref{eq-D6}]. For a given $x$,
\begin{align}
    g(x)=\int_{0}^{x}xdx+x\int_{x}^{1}dx=x-\frac{x^2}{2}.
\end{align}
Note that the average normalized aggregate output of the leaf nodes $l\in\mathcal{S}(i)$ supplying node $i$ is:
\begin{align}
    \mathbb{E}[z^dI_l^{(d)}]=\lim_{z\to\infty}\frac{1}{z}\sum_{l\in\mathcal{S}(i)}z^dm_l^{(d)}=1/2.
    \label{eq-D10}
\end{align}
For example, for a regular tree with height $h=2$, we can check that $\mathbb{E}[I_i^{(0)}]=g(1/2)=3/8$ by using Eq.~\eqref{eq-D6} and~\eqref{eq-D10}, and the result is consistent with Eq.~\eqref{eq28} in the main text.

We can extend this rule for a general height $h$. As a result, critical demand in the infinite-branching limit, $R_h := \lim_{z\to\infty} r^*(h, z)$, has the following nonlinear recurrence relation:
\begin{align}
    R_{h+1}=R_h-\frac{R_h^2}{2},\quad R_1=\frac{1}{2}.
\end{align}
Alternatively,
\begin{align}
    R_h=\underbrace{g(g(\cdots g}_{h}(1))),
\end{align}
where $g(1)=1/2$. That is, $R_h$ can be expressed by applying $g(x)$ recursively $h$ times.

Applying Eq.~\eqref{eq-D8} to Eq.~\eqref{eq-D2}, we can express the root's average production as
\begin{align}
    g(y)=y-\frac{y^2}{2},\quad0\leq y\leq1,
\end{align}
where $y=\sum_{l\in\mathcal{S}(i)}I_0^{(1)}$ is the aggregate input of the root, $g(y)=\mathbb{E}[I_0|Y=y]$ is the average production of the root for given $y$, and $\mathbb{E}[g(Y)]=r^*$ is critical demand. The function $g(y)$ describes the nonlinear response to $y$ and is concave, since its second derivative is $g''(y)=-1<0$. Jensen's inequality therefore gives:
\begin{align}
    \mathbb{E}[g(Y)]\leq g(\mathbb{E}[Y])\longrightarrow r^*\leq g(\mathbb{E}[Y]).
\end{align}
In other words, even when average production capacity remains unchanged, stochastic fluctuations in supply reduce average output relative to a scenario in which aggregate supply is constant at its average. This is the ``fluctuation-induced production loss'' that we mentioned in the main text.

For $h=2$, this loss can be calculated precisely:
\begin{align}
    g(\mathbb{E}[Y^{(w)}])-\mathbb{E}[g(Y^{(w)})]=\frac{1}{2}{\rm Var}(Y^{(w)}).
\end{align}
This is equivalent to $1/(24w)$ in Eq.~\eqref{eq28}.

\section{\label{appendix-E} Critical demand in a regular lattice\\ for fully-connected and infinite-width limit}

We consider a substitutable supply network with no stock capacity $(s=0)$. As in Appendix~\ref{appendix-D}, we start with Eq.~\eqref{eq-D1}.

In this paper, we refer to a directed regular lattice without the root node as a {\it regular lattice}. In this case, every node that is neither a root nor a leaf has in-degree and out-degree $K$; that is, $|\mathcal{S}(i)|=|\mathcal{C}(i)|=K$.

We consider the case in which the degree $K$ equals the lattice width $W$, i.e., $K=W$. Under demand rationing, each node except the root has demand $D_i^{(d)}=D_0/W$. Accordingly, the root node places orders $\Omega_{0l}^{(0)}=D_0/W$, while all other nodes place orders $\Omega_{il}^{(d)}=D_i^{(d)}/W$. Applying this to delivery rule, the root node receives $a_{0l}^{(0)}=I_l^{(1)}$, whereas any other node $i$ receives $a_{il}^{(d)}=\frac{D_i^{(d)}/W}{D_l^{(d+1)}}I_l^{(d+1)}$. Thus,
\begin{align}
    I_i^{(d)}=
    \begin{cases}
    m_i^{(d)}
    & \text{if $i$ is leaf,} \\
    \min[m_i^{(d)},\sum_{l\in\mathcal{S}(i)}I_l^{(d+1)}]
    & \text{if $i$ is root,} \\
    \min[m_i^{(d)},(1/W)\sum_{l\in\mathcal{S}(i)}I_l^{(d+1)}]
    & \text{otherwise.}
    \end{cases}
\end{align}
Here $I_i^{(d)}$ and $m_i^{(d)}$ are the output and production capacity of node $i$ at depth $d$. According to Eq.~\eqref{eq23} and~\eqref{eq24}, each production capacity $m_i^{(d)}$ is given by:
\begin{align}
    m_i^{(d)}\sim
    \begin{cases}
    {\rm Uniform}(0,1)
    &\text{if $i$ is root,} \\
    {\rm Uniform}(0,1/W)
    &\text{otherwise.}
    \end{cases}
    \label{eq-E1}
\end{align}
The root node output can be rewritten as follows:
\begin{align}
    I_i^{(0)}=\min\left[m_i^{(0)},\frac{1}{W}\sum_{l\in\mathcal{S}(i)}WI_l^{(1)}\right],
    \label{eq-E3}
\end{align}
where $m_i^{(0)}=U_i^{(0)}$ is the standard uniform random variable. Multiplying $W$ to both sides of Eq.~\eqref{eq-E1}, the non-root node output satisfies the below equation:
\begin{align}
    WI_i^{(d)}=\min\left[Wm_i^{(d)},\frac{1}{W}\sum_{l\in\mathcal{S}(i)}WI_l^{(d+1)}\right],
    \label{eq-E4}
\end{align}
where $Wm_i^{(d)}=U_i^{(d)}$. For the infinite-width limit $W\to\infty$, the law of large numbers gives:
\begin{align}
    \frac{1}{W}\sum_{l\in\mathcal{S}(i)}WI_l^{(d)}\xrightarrow{W\to\infty}\mathbb{E}[WI_l^{(d)}].
    \label{eq-E5}
\end{align}
By applying Eqs.~\eqref{eq-E4} and~\eqref{eq-E5} to Eq.~\eqref{eq-E3}, we obtain the average production of the root node as follows:
\begin{align}
    \mathbb{E}[I_i^{(0)}]=\mathbb{E}[ \min( U_i^{(0)}, \mathbb{E}[  \min(U_j^{(1)}, \dots)] ) ].
\end{align}
This result is equivalent to critical demand in a regular tree for the infinite-branching limit in Eq.~\eqref{eq32}. Equivalently,
\begin{align}
    \lim_{W\to\infty}r_{\rm lattice}^*(h,W,K=W)=\lim_{z\to\infty}r_{\rm tree}^*(h,z).
    \label{eq-E7}
\end{align}
We can see that numerical simulation results in Fig.~\ref{fig6} are consistent with Eq.~\eqref{eq-E7}.

\section{\label{appendix-F} Complementary numerical simulations}

\subsection{Effect of monopolistic layers}
\begin{figure}[b]
\includegraphics[width=0.95\columnwidth]{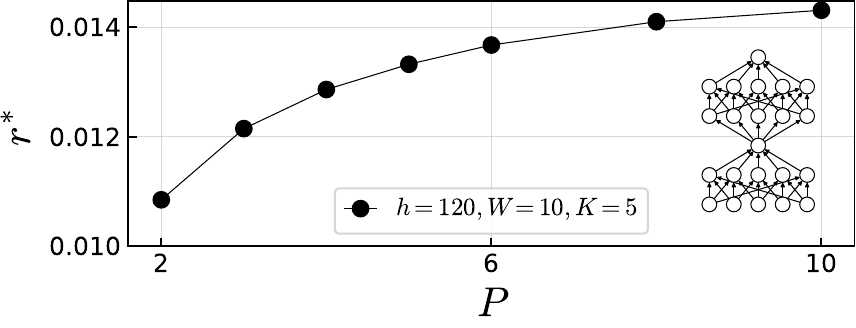}
    \caption{Critical demand $r^*$ against the period of the monopolistic layer $P$ in a repeated monopoly structure with height $h=120$, width $W=10$, and connectivity $K=5$. Symbols represent numerical simulation results. The inset shows a schematic illustration of the network structure.}
    \label{figA1}
\end{figure}
Some supply networks contain unavoidable intermediate nodes, such as logistics hubs, ports, or monopolistic suppliers. In network terms, these nodes have high betweenness centrality~\cite{goh2001universal} and can create bottlenecks even when inputs are otherwise substitutable. To examine this effect, we consider a repeated monopoly structure in which monopolistic layers interrupt the regular lattice at intervals of $P$ [see the inset in Fig.~\ref{figA1}].

Figure~\ref{figA1} shows critical demand $r^*$ as a function of the period $P$ between monopolistic layers, with fixed height $h=120$ and fixed width $W=10$. More frequent monopolistic layers reduce critical demand, indicating that more frequent bottlenecks paralyze supply network performance even when $h$ is fixed.

\subsection{\label{stock_capacity} Effect of stock capacity}

In the main text, we focused on the case with no stock capacity, $s=0$, to isolate the effects of production capacity volatility and input substitutability. Here we examine how stock capacity modifies these effects.

To briefly analyze the stock capacity effect, we refer to Eqs. \eqref{eq19} and \eqref{eq20} in the main text. For convenience, we consider a substitutable supply network case. When stock is present, the production function is given by:
\begin{align}
    I_i(t)=\min[m_i(t),D_i(t),A_i(t)+k_i(t)],
\end{align}
where $A_i(t)=\sum_{l\in\mathcal{S}(i)}a_{il}(t)$ is the aggregate supply. The stock is updated as follows:
\begin{align}
    k_i(t+1)=\min[s,k_i(t)+A_i(t)-I_i(t)].
\end{align}
Here $A_i(t)+k_i(t)$ is the sum of the inputs supplied and the existing stock at time $t$. Some of this supply is used for production, while the remainder is carried over to the next period as stock. Any amount that exceeds the stock capacity $s$ cannot be stored and is lost. We define the inputs that cannot be stored as:
\begin{align}
    Q_i(t):=\max[0,k_i(t)+A_i(t)-I_i(t)-s].
\end{align}
Consequently, the following conservation equation holds:
\begin{align}
    A_i(t)+k_i(t)=I_i(t)+k_i(t+1)+Q_i(t).
\end{align}
Summing over a period of length $T$, we obtain:
\begin{align}
    \sum_{t=1}^{T}I_i(t)=\sum_{t=1}^{T}A_i(t)+k_i(1)-k_i(T+1)-\sum_{t=1}^{T}Q_i(t).
\end{align}
In a stationary state with finite stock capacity $s$, the long-term average eliminates the terms involving the initial and final stock values. Therefore,
\begin{align}
    \langle{I_i}\rangle=\langle{A_i}\rangle-\langle{Q_i}\rangle.
\end{align}
Similarly, for a non-substitutable supply network,
\begin{align}
    \langle{I_i}\rangle=\langle{a_{il}}\rangle-\langle{Q_{il}}\rangle\quad{\forall l\in\mathcal{S}(i)},
\end{align}
where $Q_{il}(t):=\max[0,k_{il}(t)+a_{il}(t)-I_i(t)-s]$. These equations clarifies the role of stock capacity. Stock capacity does not create additional production capacity. Instead, it allows unused inputs to be carried over to subsequent periods, potentially reducing overflow losses caused by temporal mismatches between supply and production~\cite{graves2000optimizing}. This is a temporal-buffering mechanism that resolves mismatches across time, distinct from the spatial-averaging mechanism discussed in Appendix~\ref{appendix-D}, which mitigates fluctuation-induced production loss.

\begin{figure}[t]
\includegraphics[width=0.95\columnwidth]{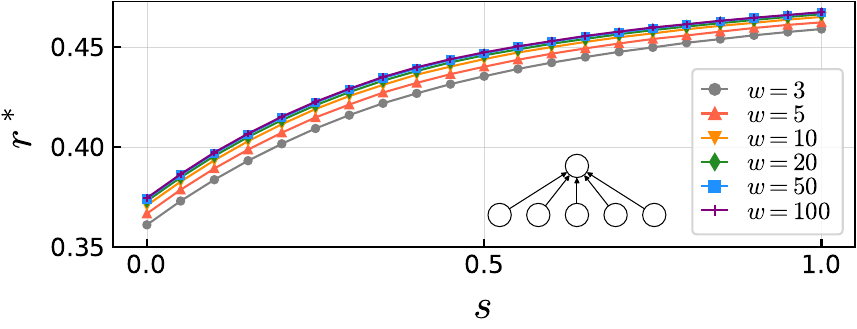}
    \caption{Critical demand $r^*$ against stock capacity $s$ in a multiple-chain with height $h=2$ for various widths $w$. Each symbol represents numerical simulation results for $2000$ time steps. The inset shows a schematic illustration of the network structure.}
    \label{figA2}
\end{figure}

\begin{figure}[t]
\includegraphics[width=\columnwidth]{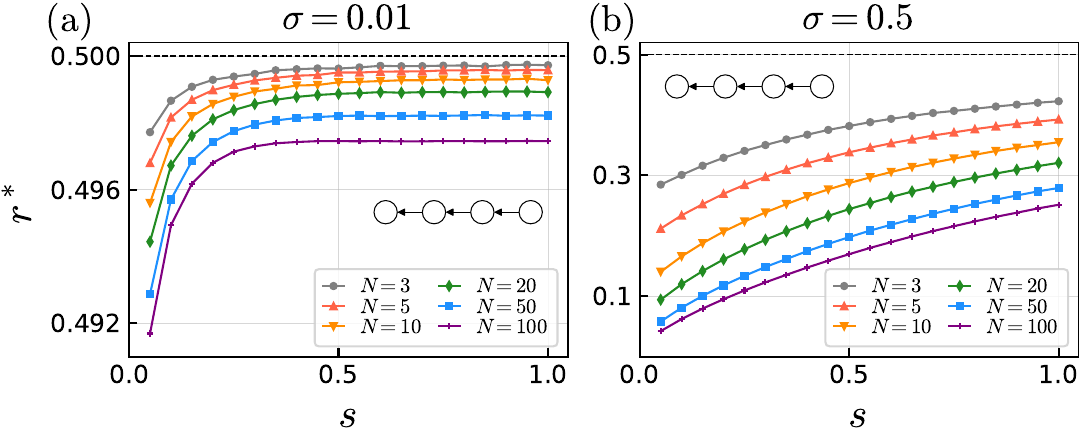}
    \caption{Critical demand $r^*$ against stock capacity $s$ in a single chain for truncated normal-distributed capacities with $\mu=1/2$ and (a) $\sigma=0.01$ and (b) $\sigma=0.5$. Each symbol represents numerical simulation results for $2000$ time steps. The insets show schematic illustrations of the network structure.}
    \label{figA3}
\end{figure}

Figures~\ref{figA2} and~\ref{figA3} consistently show that, for both substitutable and non-substitutable supply networks, critical demand $r^*$ increases with stock capacity $s$. However, as with volatility and substitutability, the effect of stock capacity is also bounded: critical demand cannot exceed the root node’s average production capacity, $\mathbb{E}[m_0]=1/2$, regardless of how large the stock capacity is.

An important finding is that stock capacity $s$ interacts with capacity volatility $\sigma$. Stock $k(t)$ primarily buffers temporal mismatches between delivered inputs $a_{il}(t)$ and actual production $I_i(t)$ [see Eqs.~\eqref{eq6} and~\eqref{eq20}]. When volatility is low, production and delivery are nearly deterministic; only small mismatches between supply and production result in little stock accumulation, and the effect of stock capacity is restricted. Consequently, critical demand $r^*$ shows early saturation for $\sigma=0.01$ in Fig.~\ref{figA3}(a), but does not saturate for $\sigma=0.5$ in Fig.~\ref{figA3}(b) or for ${\rm Var}(m_0)=1/12$ in Fig.~\ref{figA2}.

\begin{figure}[b]
\includegraphics[width=\columnwidth]{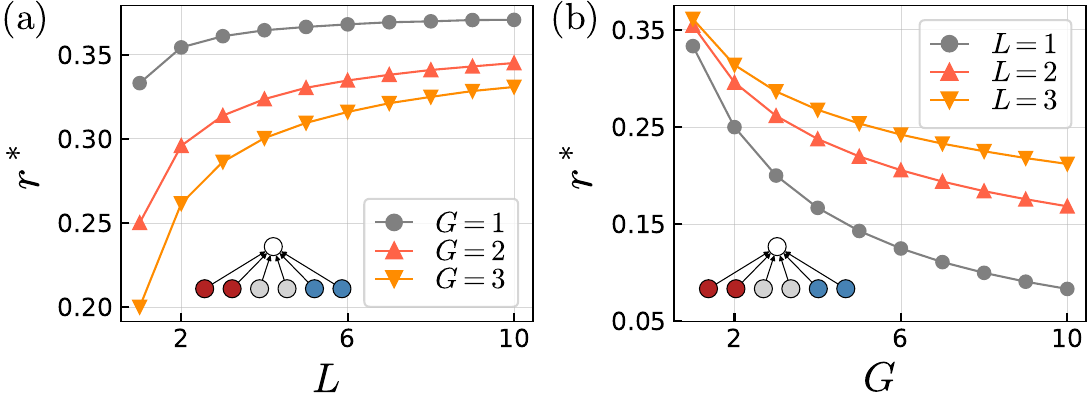}
    \caption{Critical demand $r^*$ against (a) the number of links per input type $L$ and (b) the number of input types $G$ in the case of network height $h=2$. Each symbol represents numerical simulation results for no stock capacity $s=0$ and uniformly distributed production capacities. The insets show a schematic illustration of a hybrid supply network, and the different colors represent different types of producers.}
    \label{figA4}
\end{figure}

\vspace{0.5cm}
\subsection{Effect of mixed suppliers}

The main text considers cases involving either only non-substitutable inputs or only substitutable inputs. However, in real-world supply chains, firms typically obtain multiple types of non-substitutable inputs from multiple suppliers~\cite{pichler2022forecasting,barrot2016input}. A simple extension of the production function can represent this more general situation:
\begin{align}
    I_i(t)=\min\left[m_i(t),D_i(t),\min_g\left(\sum_{l\in\mathcal{S}_{ig}}a_{il}(t)+k_{ig}(t)\right)\right].
\end{align}
Here $g$ is the type of input used by firm $i$, $\mathcal{S}_{ig}$ is the set of suppliers $l$ supplying input type $g$ to firm $i$, and $k_{ig}(t)$ is the stock of firm $i$ for input type $g$.

In the main text, we showed that the more non-substitutable inputs (e.g., network height $h$), the lower critical demand, whereas the more substitutable inputs (e.g., network width $w$), the higher critical demand. Thus, increasing the number of input types $G$ is expected to lower critical demand, while increasing the number of suppliers $L$ for a given input type $g$ is expected to raise it. Figure~\ref{figA4} shows numerical simulation results for critical demand $r^*$ in a hybrid supply chain with height $h=2$, against $L$ and $G$. The results are consistent with the intuition developed in the main text.

\end{appendix}

\bibliographystyle{apsrev4-2}
\bibliography{ref}

\end{document}